# A spacetime ontology compatible with quantum mechanics

Roderick I. Sutherland

Centre for Time, University of Sydney, NSW 2006 Australia

rod.sutherland@sydney.edu.au

The ontology proposed in this paper is aimed at demonstrating that it is possible to understand the counter-intuitive predictions of quantum mechanics while still retaining much of the framework underlying classical physics, the implication being that it is better to avoid wandering into unnecessarily speculative realms without the support of conclusive evidence. In particular, it is argued that it is possible to interpret quantum mechanics as simply describing an external world consisting of familiar physical entities (e.g., particles or fields) residing in classical 3-dimensional space (not configuration space) with Lorentz covariance maintained.

## 1. Introduction

In looking at the various possible interpretations of quantum mechanics which have been proposed over the years, one is struck by the extent to which most introduce ontology that is radically different from classical mechanics. Examples one might consider are the drastically expanded reality required in the many worlds interpretation and, more generally, the relegation of physical reality to 3n-dimensional wavefunctions residing in configuration space. Pondering such far-reaching conceptual shifts raises the question of how many of the well-known classical features could actually be maintained without losing consistency with the quantum mechanical predictions if one were to attempt a more minimalist change. Examples of such features which have served well in the past are the 4-dimensional picture of special relativity (leading to the elegant description of gravitational phenomena in terms of curved spacetime), the equal status of all inertial reference frames, the strict conservation of momentum, energy and angular momentum arising from various pleasing symmetries and the representation of reality in terms of entities interacting smoothly without discontinuous changes. The aim here is to demonstrate that an interpretation involving minimal alteration of these concepts is possible and that it can be constructed in a straightforward manner.

Of course, going down this path already assumes implicitly that some ontology does, in fact, exist and that its precise nature should be pursued. In contrast, there is always the basic objection that any events occurring at times between measurements are, by definition, beyond the reach of observation and so could be considered to have no place in physics (being thereby banished to metaphysical discussions). Nevertheless, this does not stop many physicists from talking confidently about what exists. For example, it is a common view that quantum entities necessarily become fuzzy and indefinite between measurements (even though all measurement results are found to be quite definite). In the context of quantum field theory, it is often said that the world consists only of fields (despite the fact that most experimental outcomes are seen to be localised, "particle-like" events). Furthermore, in the context of the path integral approach it is sometimes said that quantum particles somehow go along all possible trajectories, not just one. The truth is that we actually have no idea what exists in the absence of observation, despite these confident, but contradictory, pronouncements.

The diversity of opinion on the interpretation issue basically stems from the fact that quantum mechanics has proved to be no more than a mathematical prescription (albeit a highly successful one) which predicts the outcomes of measurements without giving a clear picture of a physical reality existing continuously at other times. While the formalism of classical mechanics makes the nature of its underlying entities seem self-evident, the quantum formalism is not so unambiguously interpretable. In spite of the complete lack of consensus so far, the present author believes we must try to make progress on this problem because otherwise we may deprive ourselves of an important clue towards future directions in physics. In particular, contemplation of apparently unmeasurable things may point the way to new things which are measurable (thereby indirectly verifying the original conjectures). Given that there are insufficient footholds to guide us in this endeavour, however, it would seem a plausible option to pursue a cautious approach by trying to keep as many as possible of the pre-quantum concepts which have not been excluded by subsequent experimental evidence (as opposed to exclusion by philosophical preference) and then to formulate a theory based on these notions.

Such a plan will be followed here by starting with the existing formalism of quantum mechanics and combining it with a consistent "add-on", namely a mathematical description of a proposed underlying physical reality. Maintaining the standard formalism means there will automatically be no contradiction with the usual predictions of quantum mechanics, while keeping open the possibility that the resulting model could make extra predictions in regions where standard quantum mechanics cannot.

**2. Assumptions**

Below is a summary of the assumptions on which the suggested ontological picture will be based, these being well verified and accepted in pre-quantum physics and not ruled out by any later evidence. In addition to the obvious assumption of realism[1], i.e., that there exists such an underlying physical picture to be found, the following assumptions are made:

1. Spacetime:

> It is assumed that all physical entities reside within the 3+1 dimensional spacetime of relativity. In particular, physical reality is not relegated to 3n-dimensional configuration space when n entangled particles are described by a single, overall wavefunction.

---

[1] By "realism" the present author means the view that there is an external physical reality which exists in the absence of observation.

2. Lorentz covariance:

   It is assumed that this symmetry remains valid at the hidden level as well as at the observable level, so that there is no preferred reference frame. This assumption greatly restricts the possible models available. Justification for this assumption comes not only from the fact that there is no experimental evidence whatsoever contradicting relativity, but also from the pleasing theoretical simplifications which relativity introduces by uniting various previously separate concepts. For instance, not only do space and time become more closely related, but also mass and energy, electric and magnetic fields, energy and momentum, etc.

3. Continuity:

   It is assumed that all processes occur in a continuous manner. For example, it is assumed that the ontological state at a time slightly before or after a measurement is only slightly different from the state at the time of measurement, with sudden, discontinuous changes being excluded. This assumption provides a means of interpolating between measurements to obtain a consistent picture of the intervening reality. Note that the alternative possibility of discontinuous change is not easily amenable to mathematical description nor physical explanation and is prone to violate Lorentz covariance.

4. Lagrangian:

   It is assumed that a Lagrangian formulation is possible. This assumption provides a mathematical starting point and allows the model to mesh easily with other parts of physics. It also provides a formalism which is sufficiently rich to answer any question we wish to ask. Although we should not necessarily expect this feature of classical physics to be applicable in the quantum realm and so should be wary of including superfluous formalism, there is positive evidence in favour of this assumption, namely that Lagrangian formulations have continued to prove applicable and useful in non-classical areas of physics such as quantum field theory and the standard model, where they are used to derive the propagators and vertex factors for Feynman diagrams.

5. Universality:

   It is assumed that the basic laws of quantum physics apply universally, no special status being given to observers, measurement interactions or the macroscopic world. This assumption asserts that there is no need for mind or consciousness to be involved in order to cause an instantaneous collapse of the wavefunction, nor is there a need for such a collapse in order to solve the measurement problem. It also asserts that there is no need to invoke a separate classical world outside of quantum mechanics in order to describe and understand quantum phenomena.

Having stated the above assumptions, the question now arises as to what sort of continuously-existing entities are to be described. The most obvious choices would be that they are either particles or fields, i.e., either localised entities or spread-out entities, these being the two opposing possibilities to consider once space and time have been introduced via assumption 1. The present author does not actually have a firm commitment to either option. Indeed, the mathematics to be outlined below can be employed in the context of either of these two apparently different pictures. The reason is as follows. In order for a "particle reality" to be consistent with standard quantum mechanics, the motion of each particle must be non-uniform. The proposed particles could not just move with constant speed and constant direction in free space because that would correspond to classical rather than quantum mechanics. From a mathematical viewpoint, the non-uniform motion can most conveniently be described in terms of a hypothetical field influencing each particle, so that a field effectively enters the particle picture too[2]. It is then found, however, that the specific field formalism derived for this task can also be taken as describing a "field reality"[1], with the particles then able to be deleted as superfluous! The same mathematics therefore provides simple examples of both a particle-type model and a field-type model consistent with the previously stated assumptions. It should be emphasised here that those assumptions taken together are not easily satisfied (particularly not assumption 1 that the fields must continue to be defined in 3-dimensional space rather than configuration space).

Having outlined these preliminary considerations, an ontological choice will now be made.

6. Both particles and fields

> It is assumed that quantum mechanics is describing continuously existing particles having world lines in spacetime, each particle being influenced by an associated field[3]. The main reason for the particle choice is that the existence of such localised entities provides easy and immediate agreement with the localisation observed in experiments. For example, it would seem that a particle trajectory is a more natural explanation for cloud chamber tracks than appealing to repeated wavefunction collapses. Also, the sudden localised spot made by an electron on a simple photographic plate is easier to understand if the electron is already a localised object before striking the plate.

Further support for a particle description can be obtained by considering the measurement problem of quantum mechanics. Once the existence of a hidden particle trajectory is assumed in addition to the wavefunction formalism, one can take advantage of the well-known Bohm theory of measurement [2] to provide a simple explanation for why a definite value is always found for any measured quantity, in conflict with the predicted continuous Schrodinger evolution. The explanation is that any measurement interaction must always spatially separate the possible outcomes in some way so that we can distinguish between them and identify the result. The moving particle can then flow into only one of the wavefunction branches which have become separated in this manner, thereby rendering the other branches irrelevant from then on. So-called wavefunction collapse is then seen to be simply the decision to ignore the empty branches in so far as they will have no further physical relevance. This alternative explanation for wavefunction reduction demonstrates why it is unnecessary to invoke the intervention of a conscious observer for this purpose.

---

[2] The inclusion of this field in the picture then provides a simple explanation for interference phenomena as well.
[3] The connection between these fields and the standard wavefunctions will be clarified in later sections.

The introduction of particles into the picture requires the following more specific form of Assumption 3 (continuity) to be stated:

3. Continuous and smooth world lines:

> It is assumed that the trajectories of the proposed particles are unbroken (i.e., no gaps). This implies that the equation of continuity holds in spacetime and can be employed in the model. It is further assumed that there are no discontinuous changes in the velocities of the particles, so there are no sharp corners in the world lines. This allows the particles to be described by differential equations of motion.

The second part of this assumption (concerning smoothness) is more contentious, since stochastic models of quantum mechanics assume discontinuous changes of direction and Feynman diagrams of electrons are seen to have sharp zig zags. Nevertheless, the following can be argued in favour of smooth trajectories. If a particle in free space suddenly changes its speed or direction discontinuously, the question arises as to why it did so at that location in spacetime, rather than at some other point. In the absence of a cause, this would violate the homogeneous symmetries of space and time. If, on the other hand, a cause is assumed, this would mean introducing some extra structure (e.g., a contrived underlying medium to provide collisions) for which there is no other evidence nor need.

On the basis of the above assumptions, a model will now be outlined in which the statistical nature of quantum mechanics arises because of our unavoidable lack of knowledge of each particle's position and velocity.

**3. Proposed model satisfying the chosen assumptions**

In demonstrating a viable mathematical model, the single-particle case will be considered first, followed by the generalisation to more than one particle. Despite an extensive search, the present author has only been able to find one model which satisfies all the assumptions listed above. This model has been formulated in detail in [3] and [4] and only the basic framework will be outlined here. In accordance with assumption 4, the essential idea is to introduce a Lagrangian density expression describing the behaviour of each underlying particle, as well as of any associated field. All the relevant equations for the model then flow from this single expression. In particular, the particle equation of motion and the field equation are both derivable via variational techniques. In addition, the symmetries of the expression automatically guarantee conservation of energy and momentum in the mutual interaction between particle and field.

Now, by analogy with the well-known case of classical electromagnetism, the Lagrangian density $\mathcal{L}$ is taken to have the following form:

$$\mathcal{L} = \mathcal{L}_{\text{field}} + \mathcal{L}_{\text{particle}} \tag{1}$$

The aim is to formulate a general expression for $\mathcal{L}$ which is applicable for any relativistic wave equation, in particular for both the Dirac and Klein-Gordon equations. For the single-particle case, the effective field guiding the motion of the particle can simply be identified with the wavefunction of the relevant wave equation. Fortunately, a Lagrangian density expression for each wave equation is readily available in text books and can simply be inserted here for the term $\mathcal{L}_{\text{field}}$ in Eq. (1). It therefore only remains to identify an appropriate form for the term $\mathcal{L}_{\text{particle}}$. Again, it is useful at this point to look at the analogous expression used in classical electromagnetism. For a particle of 4-velocity $u^\alpha$ under the influence of a 4-potential $A^\alpha$, this expression has the following manifestly Lorentz covariant form[4]:

$$\mathcal{L}_{\text{particle}} = \sigma_0 [-m(u_\alpha u^\alpha)^{1/2} - q u_\alpha A^\alpha] \qquad (2)$$

It is not difficult to show that a similar expression can satisfy the present requirements. The transition to the desired expression is made by recalling that there is a 4-vector associated with each wave equation in quantum mechanics, namely the conserved 4-current density $j^\alpha$. Using this 4-vector, which is always a bilinear function of the corresponding wavefunction $\psi$, an appropriate expression for the particle term $\mathcal{L}_{\text{particle}}$ can be obtained from Eq. (2) simply by replacing $-A^\alpha$ with the relevant expression for $j^\alpha$ and replacing m with the magnitude of this $j^\alpha$. This yields the following general result for the quantum mechanical case[5]:

$$\mathcal{L}_{\text{particle}} = \sigma_0 [-\rho_0 (u_\alpha u^\alpha)^{1/2} + u_\alpha j^\alpha] \qquad (3)$$

where the symbol $\rho_0$ has be used here to represent the magnitude of the 4-vector $j^\alpha$.

It is now a straightforward matter to verify consistency with quantum mechanics by inserting this expression back into the overall Lagrangian density $\mathcal{L}$ in Eq. (1) and using standard Lagrangian techniques to derive all the relevant equations of the model. The resulting formalism is actually found to be more general than is required in order to coincide with quantum mechanics and an extra assumption is needed to limit it appropriately. This last point can be illustrated by looking at the field equation which follows from the chosen Lagrangian density in the Dirac case. After inserting the well-known Dirac Lagrangian density into Eq. (1) to serve as $\mathcal{L}_{\text{field}}$, the following field equation can be derived[6]:

---

[4] Here the constants m and q represent mass and charge, respectively, with $\alpha$ ranging over the values 0,1,2,3 and with the convention $\hbar = c = 1$ assumed. The quantity $\sigma_0$ is the rest density distribution of the particle through space. This latter quantity will involve a delta function because the particle's "matter density" is concentrated at one point.

[5] In Eqs. (3) and (4), an arbitrary constant has been omitted for simplicity.

[6] Here $\psi$ is a spinor wavefunction and $\gamma^\alpha$ are the Dirac matrices.

$$i\gamma^\alpha \partial_\alpha \psi_i - m\psi_i = \sigma_0 \left( u_\alpha - \frac{j_\alpha}{\rho_0} \right) \gamma^\alpha \psi_i \quad (4)$$

Although the terms on the left hand side here are the familiar Dirac ones, the equation differs from the standard Dirac equation due to the new term on the right hand side. The extra assumption needed to ensure agreement with standard quantum mechanics is then simply that the particles be restricted to the trajectories specified by the well-known velocity equation of the Bohm model [2]:

$$u_\alpha = \frac{j_\alpha}{\rho_0} \quad (5)$$

This condition is seen to reduce the field equation (4) back to the familiar Dirac form. Furthermore, the same extra assumption is found to be applicable in reducing all other equations derived from the proposed Lagrangian density back to standard quantum mechanics.

**4. Many-particle case**

In considering the generalisation to more than one particle, the discussion here is limited to the case where the particles are not presently interacting, although it encompasses entanglement existing from past interactions. The main restriction to act as a guide in this generalisation is Bell's theorem [5], which focuses on the states of two entangled, but widely separated particles and concludes that any underlying reality compatible with the predicted correlations must be spatially nonlocal. More specifically, the theorem implies that we must either abandon Lorentz covariance by introducing a preferred reference frame in which the required influences between the particles propagate instantaneously, or else we must introduce backwards-in-time effects, i.e., retrocausality [6,7] into our description. Since we are committed here to maintaining Lorentz covariance, the latter path is necessary[7]. Introducing this additional feature into the model requires specification of final boundary conditions as well as the usual initial ones in order to determine a particle's hidden ontological state completely at any intermediate time. The consequence of this is that the experimenter's controllable choice of the final conditions can then exert a backwards-in-time influence on the earlier state. Unlike in classical mechanics, there is room for this to be possible in quantum mechanics because, due to the uncertainty principle, the ontological state of a particle at any time can never be fully known to us from previous measurement results.

The introduction of retrocausality has three positive consequences. First it avoids the need for any "instantaneous" transfer of information between the two particles involved in Bell's theorem. Second, it restores locality from a 4-dimensional viewpoint by providing a spacetime path via which the two particles can influence each other (namely the already-existing world lines of the particles extending backwards in time to where they previously interacted). Third, it allows a general technique [4] for reducing multi-particle wavefunctions in configuration space to separate wavefunctions in 3-dimensional space once the relevance of final boundary conditions is accepted and taken into account. From an ontological viewpoint then, the physical fields needed in spacetime in order to influence the particles in the many-particle case can be equated with the individual wavefunctions provided by this technique. Furthermore, this last feature provides an explanation for why our physical reality apparently becomes relegated to configuration space in the many-particle case, namely that we lack the necessary knowledge about the future which would allow us to identify the separate wavefunctions of the various particles. Similarly, the statistical nature of quantum mechanics can be explained in this picture as being due to our unavoidable ignorance of the final boundary conditions.

---

[7] Note that the nonlocal influence implied by Bell's theorem does not extend to the possibility of sending signals via entanglement. In keeping with this fact, the model presented here also does not allow any such signalling.

The steps needed to include final boundary conditions in the mathematical description will now be briefly outlined. One proceeds by first recalling the form taken by initial boundary conditions in quantum mechanics. These are always described by an initial wavefunction which summarises the particle's relevant past. By time symmetry, it is natural to describe the final boundary conditions via another wavefunction which summarises the particle's relevant future (e.g., the result of the next measurement to be performed). The initial and final wavefunctions introduced in this way are usually denoted by $\psi_i$ and $\psi_f$, respectively, with the latter providing the mathematical extension needed for incorporating retrocausal effects.

As shown in [4], the final wavefunction $\psi_f$ can easily be incorporated into the standard formalism while maintaining agreement with all existing predictions. As an example of this procedure, consider the expression for the 4-current density $j^\alpha$ associated with the Dirac equation:

$$j^\alpha = \overline{\psi}\gamma^\alpha\psi \tag{6}$$

As with most expressions in quantum mechanics, this expression is bilinear in the wavefunction $\psi$. With this in mind, it is found that a suitable and natural generalisation to include the final wavefunction has the form:

$$j^\alpha = \frac{1}{N}\overline{\psi}_f\gamma^\alpha\psi_i + \frac{1}{N^*}\overline{\psi}_i\gamma^\alpha\psi_f \tag{7}$$

where N is a normalisation constant. The reason this expression contains two terms is so that the two wavefunctions $\psi_i$ and $\psi_f$ can be included on an equal footing, each appearing in both "barred" and "unbarred" form. The resulting expression can be recognised as being similar in form to the "weak value" formalism of Aharonov et al [8].

In view of the above considerations, the Lagrangian formulation for the single-particle case discussed earlier can be carried over without substantial change to the entangled many-particle case simply by employing the individual wavefunctions discussed in this section. In particular, the technique defined in [4] provides the particles with individual $\psi_i$'s to complement their individual $\psi_f$'s so that the various fields all reside in spacetime. As an example, the full Lagrangian density for each particle in the Dirac case then becomes[8]:

$$\mathcal{L} = \text{Re}\left[\frac{1}{\langle f|i\rangle}\left(-i\bar{\psi}_f\gamma^\alpha\partial_\alpha\psi_i + m\bar{\psi}_f\psi_i\right)\right] + \sigma_0\left[-\rho_0(u_\alpha u^\alpha)^{1/2} + u_\alpha j^\alpha\right] \quad (8)$$

Throughout this model, the standard formalism of quantum mechanics can always be recovered simply by taking a weighted average over the unknown final boundary conditions.

Perhaps an appropriate way to finish is to give a speculative example of how choosing a suitable ontology could have implications in other areas of physics. To this end, the problem of formulating a theory of quantum gravity will be viewed in the light of the ontological model discussed here. A feature of the present model is that final boundary conditions have become relevant. The extra information contained in these final conditions then allows the model to provide a definite, non-statistical energy-momentum tensor [9] for insertion in Einstein's gravitational field equation. This, in turn, makes it viable for the Einstein curvature tensor to remain non-statistical (i.e., unquantised) as well, thereby avoiding many perplexing implications that arise from attempting to quantise space and time.

## 5. Discussion

The model outlined here demonstrates that it is possible to construct an interpretation of quantum mechanics which involves minimal alteration of the ontology that had been employed successfully in pre-quantum physics. In particular it shows that a spacetime picture can be maintained, as can Lorentz covariance. In accordance with Occam's razor there is no need to include more radical notions such as assigning ontological reality to configuration space wavefunctions, invoking a conscious observer in order to collapse the wavefunction on measurement, or elevating the macroscopic world to a status outside of quantum mechanics. There is also no requirement to have two different ways for wavefunctions to change, viz. Type 1 instantaneous collapse and Type 2 continuous evolution. Only the latter is strictly needed, as is demonstrated by the example of the Bohm theory of measurement. The model proposed here simply continues to describe an external material world, as opposed to taking a conceptual leap towards an alternative picture, such as one in which "mind" becomes primary. Finally, it is a matter of personal preference as to whether the introduction of final boundary conditions and retrocausality is seen as a welcome addition on the grounds of time symmetry, or as a price to pay in exchange for obtaining a more moderate picture.

**Acknowledgements**

I would like to thank Henry Stapp for inviting me to contribute to his 90th birthday festschrift volume. I feel it is also important to acknowledge here the commendable open-mindedness shown by Henry in inviting others to present viewpoints differing from his.

---

[8] In the more detailed presentation given in [4], a more cumbersome notation involving $\pm$ signs was employed in order to ensure that all quantities remained real. Here that notation has been dispensed with for the sake of simplicity, with the understanding that some quantities in the equations may now become imaginary in certain circumstances. This does not, of course, affect any of the observable quantities.